\documentclass[doublecol]{epl2}
\usepackage{subfigure}
\usepackage{graphicx}
\usepackage{dcolumn}
\usepackage{bm}
\usepackage{multirow}
\usepackage{caption}
\usepackage{amsmath}
\revision

\title{On Cost Effective Communication Network Designing}
\author{Guo-Qiang Zhang} \shortauthor{Guo-Qiang Zhang}

\institute{Institute of Computing Technology, Chinese Academy of
Sciences, Beijing, 100190, P. R. China }


\abstract{How to efficiently design a communication network is a
paramount task for network designing and engineering. It is,
however, not a single objective optimization process as perceived by
most previous researches, i.e., to maximize its transmission
capacity, but a multi-objective optimization process, with lowering
its cost to be another important objective. These two objectives are
often contradictive in that optimizing one objective may deteriorate
the other. After a deep investigation of the impact that network
topology, node capability scheme and routing algorithm as well as
their interplays have on the two objectives, this Letter presents a
systematic approach to achieve a cost-effective design by carefully
choosing the three designing aspects. Only when routing algorithm
and node capability scheme are elegantly chosen can BA-like
scale-free networks have the potential of achieving good tradeoff
between the two objectives. Random networks, on the other hand, have
the built-in character for a cost-effective design, especially when
other aspects cannot be determined beforehand. }

\begin{document}
\maketitle
\section{Introduction}\label{Introduction}
How to efficiently design a communication network is a subject that
involves several independent yet closely related aspects, of which
the most important three ones are what kind of network topology
shall be used, which routing algorithm is effective, and how the
node capabilities are assigned. The ultimate goal of making these
decisions is to enhance the network's transmission capacity--a
research area that has attracted substantial interest in previous
works \cite{centrality-and-network-flow, load-distribution,
role-connectivity,
optimal-network-topology,polymorphic-torus,edge-deletion,
efficient-routing, onset-traffic-congestion, Danila06,
node-capability-redistribution}, and meanwhile to lower its cost.

Previous works intending to enhance the network's transmission
capacity can be generally classified into three categories: to make
small changes to the underlying network topology
\cite{optimal-network-topology,polymorphic-torus,edge-deletion}, to
adjust the routing algorithm \cite{efficient-routing}, and to
customize the node capability \cite{onset-traffic-congestion,
node-capability-redistribution}. Although these works are
instructive, they can not be applied under arbitrary condition. For
example, the HOT model \cite{HOT} is found to be insensitive to
routing algorithm changes so that changing routing algorithms will
be ineffective. A common problem of most of these works is that they
focus only on one aspect of network designing, neglecting the fact
that network designing is a multi-objective optimization process,
and moreover, they don't consider the scalability issue.

Faced with the problem of efficiently designing a completely new
network, all the above facets and their inherent interplays should
be carefully considered. We report in this Letter a systematic
approach to design a network from the very beginning, with two
significant yet contradictive objectives: enhancing the network's
transmission capacity, and lowering its cost. Typically optimizing
one objective will deteriorate the other. The existence of a
cost-effective design depends on the particular network topology.
With elaborately chosen routing algorithm and node capability
scheme, BA-like scale-free networks can achieve pretty good tradeoff
between the two designing objectives. On the other hand, there is no
cost-effective designing for the more realistic HOT model, though it
has the same skewed degree distribution. It is striking to find that
random network is a markedly good candidate to accomplish a
cost-effective design, especially when other aspects are
undeterminable in that time. Whereas if network topology can be
determined, we strongly recommend the use of efficient routing and
effective betweenness based node capability scheme, which proves to
be cost-effective for most small-world networks. The property of
cost-effectiveness is also scalable with network size.

\section{Traffic Flow Model}\label{traffic_flow_model}

In this study, we adopt a similar traffic-flow model used in
\cite{efficient-routing, edge-deletion,
onset-traffic-congestion,Danila06}. Each node is capable of
generating, forwarding and receiving packets. At each time step, $R$
packets are generated at randomly selected sources. The destinations
are also chosen randomly. Each node is assigned a capability,
$C(i)$, which is the maximal number of packets the node can handle
at a time step. Each packet is forwarded toward its destination
based on the particular \emph{routing algorithm} used. When the
total number of arrived and newly created packets exceeds $C(i)$,
the packets are stored in the node's queue and will be processed in
the following time steps on a first-in-first-out(FIFO) basis. If
there are several candidate paths for one packet, one is chosen
randomly. Each node has a queue for receiving newly arriving
packets. Packets reaching their destinations are deleted from the
system. As in \cite{efficient-routing, edge-deletion,
onset-traffic-congestion,Danila06}, node buffer size in this
traffic-flow model is set as infinite as it is not relevant to the
\emph{occurrence} of congestion.

For small values of the packet generating rate $R$, the number of
packets on the network is small so that every packet can be
processed and delivered in time. Typically, after a short transient
time, a steady state for the traffic flow is reached where, on
average, the total numbers of packets created and delivered are
equal, resulting in a free-flow state. For larger values of $R$, the
number of packets created is more likely to exceed what the network
can process in time. In this case traffic congestion occurs. As $R$
is increased from zero, we expect to observe two phases: free flow
for small $R$ and a congested phase for large $R$ with a phase
transition from the former to the latter at the critical
packet-generating rate $R_c$.

In order to measure $R_c$, we use the order
parameter~\cite{Arenas01} $ \eta=\lim_{t\rightarrow \infty}{{\langle
\Delta\Theta \rangle} \over {R\Delta t} }$, where $\Theta(t)$ is the
total number of packets in the network at time $t$,
$\Delta\Theta=\Theta(t+\Delta t)-\Theta(t)$,  and
$\langle\cdots\rangle$ indicates the average over time windows of
$\Delta t$. For $R <R_c$ the network is in the free-flow state, then
$\Delta\Theta\approx0$ and $\eta\approx0$; and for $R>R_c$,
$\Delta\Theta$ increases with $\Delta t$ thus $\eta>0$. Therefore in
our simulation we can determine $R_c$ as the transition point where
$\eta$ deviates from zero.

\begin{table}[htbp]
\begin{center}
\caption{Elementary topological properties of the seven networks.
Ring and Lattice contain 1225 nodes, and the other five networks
contain 1200 nodes. $N$ and $M$ denote the number of nodes and
number of edges respectively, $D$ denotes the diameter, $L$ denotes
the average shortest path length. }
\label{basic_info}
\begin{tabular}{ccccc}
\hline Network & $N$ & $M$ & $D$ & $L$
\\
\hline\hline Ring & 1225 & 2450 & 306 & 153.5 \\
Lattice & 1225 & 2450 & 34 & 17.5  \\
WS & 1200 & 2400 & 15.5 & 7.86 \\
ER & 1200 & 2450 & 11 & 5.23  \\
BA & 1200 & 2390 & 8 & 4.43  \\
PA & 1200 & 2400 & 8.7 & 4.03  \\
HOT & 1200 & 2583 & 9 & 5.16  \\
  \hline
\end{tabular}
\end{center}

\end{table}

\section{Network designing objectives}\label{measuring}
\subsection{Network transmission capacity}

Enhancing network transmission capacity is one of the major goals
for network designing and engineering, and is typically measured by
the critical packet-generating rate $R_c$ \cite{efficient-routing,
edge-deletion, onset-traffic-congestion,Danila06}. Although a
simulation-based approach is feasible to measure $R_c$, it is
time-consuming. An analytical approach
\cite{optimal-network-topology} showed that, for the simplest
configuration where shortest path routing is used and each node is
assigned the same node capability $C$, the critical
packet-generating rate can be calculated by
$R_c=\frac{CN(N-1)}{B_{max}}$, where $B_{max}$ is the maximum node
betweenness centrality \cite{freeman-betweenness} value in the
network.

In our previous work \cite{comprehensive-complex-routing}, we
extended the above approach to the general case, in which, we are
provided with a network $G$, a node capability scheme that assigns
each node $i$ a capability $C(i)$, and a
\emph{topology-based}\footnote{\emph{topology-based} routing
algorithm means routing decision is made solely on the static
topological information, not on dynamic traffic information.}
routing algorithm $\Gamma$.

Similar to \cite{efficient-routing}, we introduce the effective
betweenness $B^\Gamma(i)$ to estimate the possible traffic passing
through a node under routing algorithm $\Gamma$, which is formally
defined as:
\begin{equation}
B^{\Gamma}(i)=\sum_{u\neq
v}\frac{\delta^{\Gamma}_{(i)}(u,v)}{\delta^{\Gamma}(u,v)}
\end{equation}
where $\delta^{\Gamma}(u,v)$ is the total number of candidate paths
between $u$ and $v$ under routing algorithm $\Gamma$, and
$\delta^{\Gamma}_{(i)}(u,v)$ is the number of candidate paths that
pass through $i$ between $u$ and $v$ under routing algorithm
$\Gamma$.

Following this definition, at each time step, the expected number of
packets arriving at node $i$ in free-flow state is $\frac{R
B^{\Gamma}(i)}{N(N-1)}$. In order for node $i$ not to get congested,
it follows that $\frac{R B^{\Gamma}(i)}{N(N-1)}\leq C(i)$, which
leads to $R \leq \frac{C(i)N(N-1)}{B^{\Gamma}(i)}$. So the critical
packet-generating rate $R_c$ is:
\begin{equation}\label{fomular}
R_c=min_i \frac{C(i)N(N-1)}{B^{\Gamma}(i)}
\end{equation}

\begin{table*}[htb]
\begin{center}
\caption{Theoretical result of the critical packet-generating rate
$R_c$ under different combinations for networks in Table
\ref{basic_info}. The results are obtained from Equation 2. For each
kind of network, we generate 10 instances, and the result is the
average of the 10 instances. } \centering \label{theoretical_result}
\begin{tabular}{c|ccccccc}
\hline

(node capability scheme, routing algorithm) & BA & PA & HOT & ER
&WS&
Lattice & Ring \\
\hline \hline

(UC, SPR) & 24.2& 15.5& 59.3& 157.7& 101.4& 280 & 32 \\
(UC, EFR) & 200& 116.7& 80.4&346.9 & 144.8& 280 & 32 \\
\hline
 (DC, SPR) & 322.1 & 401.7& 177.3 & 393.4 & 152.2 & 280 & 32 \\
(DC, EFR) & 286.4 & 229.8 &121.1 & 412.8 & 166.5 & 280 & 32 \\
\hline
 (BC, SPR) & 880.6 & 955.0 & 839.0 & 787.2 & 541.8& 280 & 32 \\
(BC, EFR) & 167.2 & 112. 4 & 162.6 & 353.6 & 293.3& 280 & 32 \\
\hline
 (EBC, EFR) & 660.6 & 718.4 & 773.1 & 754.2 & 538.3 & 280 & 32 \\
 \hline
\end{tabular}
\end{center}
\end{table*}

\begin{table*}[htb]
\begin{center}
\caption{Simulation result of the critical packet-generating rate
$R_c$ under different combinations for BA, PA, HOT, ER and WS
networks. For each kind of network, 10 instances are generated and
for each instance, we run ten simulations. The result here is the
average over all the simulations. } \centering
\label{simulation_result}
\begin{tabular}{c|ccccc}
\hline

(node capability scheme, routing algorithm) & BA & PA & HOT & ER &WS\\
\hline \hline

(UC, SPR) & 25.4 & 16.0  & 64.5 & 175.6   & 113.9 \\
(UC, EFR) & 219.3& 132.1 & 85.9 & 387.0 & 161.0 \\
\hline
 (DC, SPR) & 328.4 & 439.8& 187.1 & 440.0 & 168.3 \\
(DC, EFR) & 315.4 & 260.5 & 132.1 & 474.6 & 187.3 \\
\hline
 (BC, SPR) & 785.3 & 851.0 & 740.6 & 705.6 & 476.1 \\
(BC, EFR) & 191.6 & 133.1 & 185.4 & 392.4 & 325.0\\
\hline
 (EBC, EFR) & 588.1 & 634.7 & 675.5 & 676.7 & 474.8 \\
 \hline
\end{tabular}
\end{center}
\end{table*}
\subsection{Network Cost}
Lowering the cost is another goal that network designers endeavor to
accomplish. The overall cost of a network designing scheme can be
considered as the sum of individual cost of each node. Each node
differs in its node capability, and consequently its cost. Generally
speaking, larger node processing capability means higher cost. The
relationship between node cost and node capability can be abstracted
as a function $f$. Although it is unrealistic to define a specific
$f$, empirical evidence show that node cost $f(C)$ typically grows
super-linearly with node capability $C$, i.e., $f(kC)>kf(C)$. And
moreover, the property that when $k\gg1$, $f(kC)\gg kf(C)$ often
holds in reality. As a result, if we fix $\sum_i C(i)$ for the
purpose of comparing between different node capability schemes, the
total cost $\sum_i f(C(i))$ is often dominated by $f(C_{max})$,
where $C_{max}$ is the maximal node capability of a designing
scheme. In addition, besides the monetary issue, there remains the
technical feasibility of realizing a designing scheme. $C_{max}$
poses the upper bound performance requirement for a special
designing, which can be used to judge whether the corresponding
designing strategy is technically feasible with state-of-art
technologies. For these reasons, we choose $C_{max}$ to represent
the network cost.
\begin{table*}[htb]
\begin{center}
\caption{$C_{max}$ under different combinations of network
topologies, routing algorithms and node capability schemes, where
the meaning of acronyms is the same as Table
\ref{theoretical_result}.} \centering \label{max_capability}
\begin{tabular}{c|ccccccc}
\hline

(node capability scheme, routing algorithm) & BA & PA & HOT & ER &
WS &
Lattice & Ring \\
\hline \hline

(UC, *) & 4 & 4 & 4 & 4 & 4 & 4 & 4\\
(DC, *) & 60 & 118.1 & 117.4 & 12.1 & 7.0 & 4 & 4 \\
(BC, *) & 160.2 &253.9& 58.6& 20.3 & 21.9
& 4 & 4\\
(EBC, EFR)
& 13.3 & 25.4& 38.7 & 8.9 & 15.1 & 4 & 4 \\

 \hline
\end{tabular}
\end{center}
\end{table*}
\section{Results}
\subsection{Configurations}
Two topology-based routing algorithms are investigated in this
Letter: the traditional shortest path routing and the efficient
routing algorithm proposed in \cite{efficient-routing}. For a given
pair source and destination, the efficient routing algorithm chooses
a path that minimizes the sum of node degrees along the path. More
formally, the efficient routing chooses a path
$s=v_0,v_1,v_2,\cdots,v_k=t$ between $s$ and $t$ that minimizes the
objective function $\sum_{0\leq i<k}d(v_i)$, where $d(v_i)$ is the
vertex degree of $v_i$.

Each routing algorithm is applied to seven network topologies: ring,
lattice, ER \cite{erdos59, random-graphs}, WS
\cite{small-world1998}, BA \cite{BA}, PA \cite{HOT} and HOT
\cite{HOT}. The ring is constructed by placing all the nodes in a
circular ring and connecting each node to its left two nearest nodes
and right two nearest nodes. The two-dimensional lattice is
constructed in toroidal mode so that the lattice is completely
homogeneous. WS graph is built from the ring by randomly rewiring 15
percent of its edges. BA graph is constructed according to the
standard BA model with $m=2$. PA network is generated by the
following process: begin with 3 fully connected nodes, and add one
new node to the graph in successive steps, such that this new node
is connected to the existing nodes with probability proportional to
the current node degree, and finally, add some internal edges to
augment the graph by selecting both endpoints with probability
proportional to the current node degree. Finally, the HOT network is
a heuristically optimal topology for the Internet router-level
network, which can be roughly partitioned into three hierarchies:
the low degree core routers, the high degree gateway routers hanging
from the core routers, and the low degree periphery nodes connected
with the gateway routers. The HOT networks generated here follow the
same degree distributions as the corresponding PA networks. Basic
graph properties of these networks are presented in Table
\ref{basic_info}.

Four node capability schemes are considered: uniform node capability
scheme, degree based node capability scheme, betweenness based node
capability scheme and effective betweenness based node capability
scheme. In uniform node capability scheme, each node has the same
packet transmission capability. While for the other three node
capability schemes, a node's capability is proportionate to its
degree, betweenness and effective betweenness respectively. For the
purpose of comparing between different node capability schemes, we
keep the condition that the sum of node capability in a given
network remains fixed for all the four node capability schemes,
which is set to the sum of node degrees,
i.e.,$\sum_{i}{C(i)}=\sum_{i}{k_i}=2M$. Non-integer $C(i)$ is
treated in a statistical way in the simulation: at each time step,
$i$ first forwards $\lfloor C(i) \rfloor$ packets, then a random
number $r\in(0,1)$ is generated and compared against $C(i)-\lfloor
C(i) \rfloor$. If $r<C(i)-\lfloor C(i) \rfloor$, $i$ forwards
another packet.

\subsection{Network Transmission Capacity-$R_c$}
$R_c$ can be obtained by applying Equation \ref{fomular} as well as
by running the traffic flow simulation. Table
\ref{theoretical_result} presents the analytical $R_c$ values
obtained from Equation 2 for different combinations of network
topologies, node capability schemes and routing algorithms, where UC
stands for uniform capability scheme, DC stands for degree based
capability scheme, BC stands for betweenness based capability
scheme, EBC stands for effective betweenness based capability
scheme, SPR stands for shortest path routing and EFR stands for
efficient routing. Table \ref{simulation_result} reports the
simulation result. It is evident that the simulation result roughly
agrees with the theoretical analysis.


Several observations can be made from Table
\ref{theoretical_result}: (a) all networks achieve the highest $R_c$
values when shortest path routing and betweenness based node
capability scheme is used; (b) the transmission capacity of BA and
PA network is highly sensitive to both routing algorithm and node
capability scheme changes, while the transmission capacity of HOT
network is only sensitive to node capability scheme, so the only way
to improve HOT network's $R_c$ is to upgrade key nodes with
effective betweenness centrality values; (c) the transmission
capacities of ER and WS networks don't vibrate so much as the
heterogenous networks do; (d) efficient routing combined with
effective betweenness based capability scheme can achieve relatively
high $R_c$ values; (e) although lattice and ring are both regular
networks, their $R_c$ show drastic difference. Here we only list the
observations. The explanations as well as the consequent
inspirations will be discussed later in the Discussion section.

\subsection{Network Cost--$C_{max}$}
Table \ref{max_capability} reports the $C_{max}$ values under
different node capability schemes and routing algorithms. Comparing
it with Table \ref{theoretical_result}, we find that although (BC,
SPR) enables the largest $R_c$ values, it often incurs high
$C_{max}$ values, most often the highest among all combinations. But
for (EBC, EFR), we observe significant decrease of $C_{max}$,
especially for BA and PA networks, while the $R_c$ is only slightly
lower than the optimal one. Also, we will elaborate more on this in
the Discussion section.

\begin{table}

 \caption{$R_c$ and $C_{max}$ in (UC,SPR), (UC,EFR), (BC,SPR) and (EBC,EFR)}
 \label{performance_gain}
  \centering
 \begin{tabular}{c||c|c}
 \hline
  & $R_c$ & $C_{max}$ \\
  \hline
  \hline
(UC, SPR) & $\frac{4N(N-1)}{B_{max}}$ &4\\
\hline
(UC, EFR) & $\frac{4N(N-1)}{B^{(EFR)}_{max}}$& 4 \\
\hline
(BC, SPR) & $\frac{4N}{L_G+1}$& $\frac{4B_{max}}{(N-1)(L_G+1)}$ \\
\hline
(EBC, EFR) &$\frac{4N}{L^{(EFR)}_G+1}$ &$\frac{4B^{(EFR)}_{max}}{(N-1)(L^{(EFR)}_G+1)}$ \\
\hline
\end{tabular}
\end{table}

\section{Discussion}
\subsection{Impact of different designing factors}
 We will investigate the influence of the three designing
factors on the two disparate designing objectives in this section.
Among different combinations of routing algorithms and node
capability schemes, we are primarily interested in (UC,SPR),
(UC,EFR), (BC,SPR) and (EBC,EFR).

Shortest path routing is widely used in communication networks.
Within this framework, uniform node capability scheme ensures lowest
$C_{max}$, but may at the cost of sacrificing $R_c$, especially for
networks with heterogenous structures. On the other hand,
betweenness based node capability scheme can achieve optimal $R_c$,
but typically requests high $C_{max}$, again particularly true for
heterogenous networks. In this sense, (UC, SPR) and (BC, SPR) can be
regarded as two extremes in the designing space for any given
network, as each optimizes one objective. One way to enhance $R_c$
while keeping the lowest $C_{max}$ is by applying the efficient
routing algorithm, which is proven to be effective in BA network by
Yan \cite{efficient-routing}. We have confirmed in this paper that
efficient routing is indeed effective in BA and PA, which are
flexible in path selection so that the chosen path can bypass those
high degree nodes. However, we show that efficient routing is not
effective in HOT network, which, although has the same degree
distribution as PA, exhibits more rigid hierarchical structure, so
paths are unlikely to bypass the core nodes. Also, efficient routing
is not effective in WS network because of the nearly identical
degrees. The scheme (EBC, EFR) that we proposed in this Letter can
accomplish a cost-effective design in most cases.
\begin{figure}[htb]
\includegraphics[width=8cm]{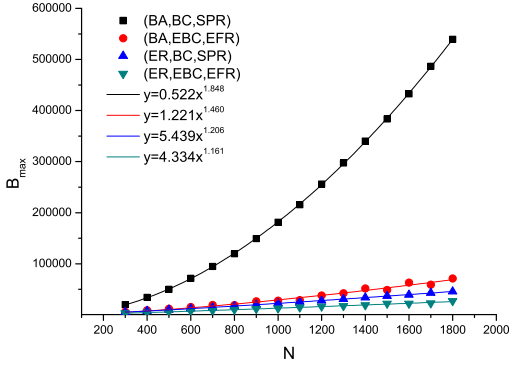}
 \caption{(Color online) Scalability of the maximal betweenness and
effective betweenness for BA and ER networks.  With SPR, the
$Y$-axis denotes the value of maximal betweenness, and with EFR, the
$Y$-axis denotes the maximal effective betweenness. The line
fittings for these data are also presented. }\label{bmax}
\end{figure}

\begin{figure}[htb]
\includegraphics[width=8cm]{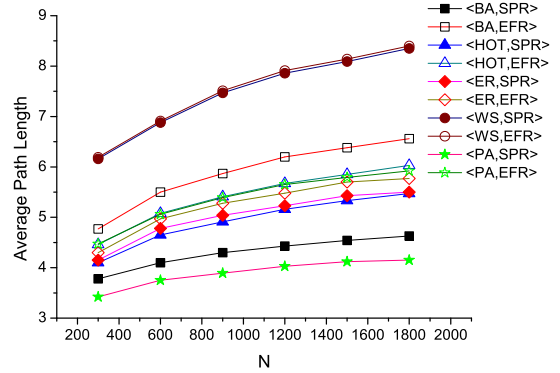}
\caption{(Color online) Average path length for the five networks
with different size under both shortest path routing and efficient
routing. The filled shape represents the SPR, while the hollow shape
represents the EFR. } \label{apl}
\end{figure}
The performance gain or tradeoff are indeed quantifiable in these
four schemes. Table \ref{performance_gain} presents the $R_c$ and
$C_{max}$ of these four schemes. Recall that with uniform node
capability scheme, $R^{\Gamma}_c=\frac{CN(N-1)}{B^{\Gamma}_{max}}$,
so the result for (UC, SPR) and (UC, EFR) is obvious.

For $R_c$ in (BC, SPR), we rewrite Equation 2 as
follows\footnote{Note that when defining betweenness, we assume each
node is on its path to all other nodes, so the betweenness of node
with degree one is 2(N-1), rather than 0, which otherwise would
incur zero node capability. Following this definition,
$\sum_{j=1}^{N}{B(j)}=N(N-1)(L_G+1)$, rather than
$\sum_{j=1}^{N}{B(j)}=N(N-1)L_G$.}:
\begin{equation}
\begin{split}
R_c& =min_i \frac{C(i)N(N-1)}{B(i)} \\
&= min_i \frac{B(i)/\sum_{j=1}^{N}B(j)\times 4N\times N(N-1)}{B(i)} \\
&=\frac{4N^2(N-1)}{\sum_{j=1}^{N}B(j)}\\
&=\frac{4N^2(N-1)}{N(N-1)(L_G+1)}\\
&=\frac{4N}{L_G+1}
\end{split}
\end{equation}
$\quad$ where $L_G$ is the average shortest path length of $G$.

Since $\sum_{j=1}^{N}{B(j)}=N(N-1)(L_G+1)$, it is easy to
demonstrate that $C_{max}=\frac{4B_{max}}{(N-1)(L_G+1)}$. Similarly,
by replacing $B_{max}$ with $B^{(EFR)}_{max}$, and $L_G$ with
$L^{(EFR)}_G$, we get the $R_c$ and $C_{max}$ for (EBC, EFR).

From Table \ref{performance_gain}, it is easy to explain the
previous observations. In BA and PA networks, $B_{max}$ can be an
order of magnitude larger than $B^{(EFR)}_{max}$(see
Fig.\ref{bmax}), which explains why the performance gain of $R_c$ is
remarkable in these networks.

With (BC, SPR), $R_c$ is purely determined by the average path
length of the given network. Since most networks have the
small-world property, their $R_c$ values will be very large, and
will not differ significantly.  This property also explains the
drastic difference shown by ring and lattice, since their average
path lengths are $\Theta(N)$ and $\Theta(\sqrt{N})$ respectively.
However, (BC, SPR) can incur high $C_{max}$ because it is
proportional to the largest node betweeness, which can be very high
in heterogeneous networks. Indeed, as is shown in Table
\ref{max_capability}, $C_{max}$ can differ by an order of magnitude
for different networks.

For heterogenous networks, (EBC, EFR) can achieve $R_c$ slightly
lower than the optimal one while at meantime significantly reduces
$C_{max}$. This effect arises from the following facts:

\begin{equation}
\frac{R^{(BC,SPR)}_c}{R^{(EBC,EFR)}_c}=\frac{L^{(EFR)}_G+1}{L_G+1}
\end{equation}

and
\begin{equation}
\frac{C^{(BC,SPR)}_{max}}{C^{(EBC,EFR)}_{max}}=\frac{L^{(EFR)}_G+1}{L_G+1}\times\frac{B_{max}}{B^{(EBC,EFR)}_{max}}
\end{equation}

In most networks, $L^{(EFR)}_G$ is only slightly longer than $L_G$,
as is evidenced in Fig.\ref{cmax}, so the ratio between
$R^{(BC,SPR)}_c$ and $R^{(EBC,EFR)}_c$ is slightly larger than 1.
However, $C^{(BC,SPR)}_{max}$ can be several times larger than
$C^{(EBC, EFR)}_{max}$, mainly due to the potentially dramatic
difference between $B_{max}$ and $B^{(EBC,EFR)}_{max}$. This
property is the foundation of the cost-effectiveness of (EBC, EFR).

\subsection{Scalability and adaptability of $R_c$ and $C_{max}$}
\begin{figure}
\includegraphics[width=8cm]{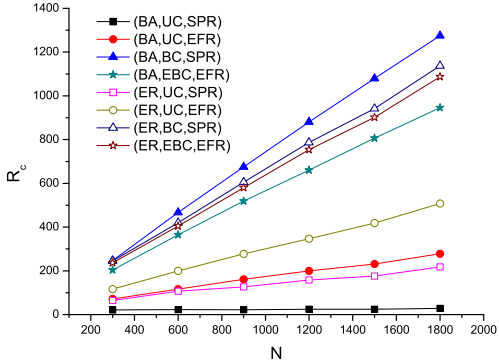}
\caption{(Color online) Scalability of $R_c$ values of BA, HOT and
ER under four different configurations. } \label{rc_evolve}
\end{figure}

\begin{figure}
\includegraphics[width=8cm]{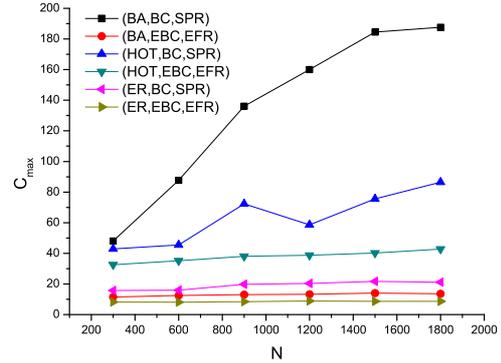}
\caption{(Color online) Scalability of $C_{max}$ values of BA and ER
under (BC,SPR) and (EBC, EFR) schemes.} \label{cmax_evolve}
\end{figure}

One question of interest is how $R_c$ and $C_{max}$ scales in
different settings as network size grows. Figure \ref{rc_evolve} and
\ref{cmax_evolve} present the $R_c$ and $C_{max}$ values for BA and
ER networks under four different settings. It is shown that under
(UC, SPR), BA network's $R_c$ value remains quite stable, i.e.,
almost not scaling with $N$. This is because, $B_{max}$ scales super
linearly with $N$, as illustrated in Fig. \ref{bmax}. Simulation
result shows that $B_{max}\sim N^{1.848}$ in BA network, whereas in
ER network, $B_{max}\sim N^{1.460}$. On the other hand,
$B^{(EFR)}_{max}$ grows much slowly as network expands, $\sim
N^{1.206}$ and $\sim N^{1.161}$ for BA and ER respectively. This
means with (UC, SPR), BA network's $R_c$ scales very slowly, while
ER network's $R_c$ scales much better. However, with (UC, EFR) the
scalability of $R_c$ significantly improves for BA network. With
(BC, SPR) or (EBC, EFR), the formula in Table \ref{performance_gain}
guarantees good scalability of $R_c$ for small-world networks.

As for $C_{max}$, with (BC, SPR), $C_{max}\propto B_{max}$, so it
grows fast for BA network, and much more slowly for ER network. With
(EBC, EFR), $C_{max}\propto B^{(EFR)}_{max}$, which scales much
slowly in BA networks. This is evidenced in Fig.\ref{cmax_evolve},
where $C_{max}$ grows fast with (BA, BC, SPR), but remains nearly
stable for (BA, EBC, EFR) and (ER, EBC, EFR). For clarity, we do not
present WS and PA in Fig. \ref{cmax_evolve}, but the growth trends
of WS and PA are similar to ER and BA respectively.


\section{Conclusion}\label{Conclusion}
In this Letter, we proposed that network designing is a
multi-objective optimization designing process and involves several
seemingly independent but in fact closely related aspects.  We found
that betweenness based capability scheme combined with shortest path
routing can achieve highest network transmission capability, but
this scheme also requires high cost. If the network topology is
predetermined and has small-world property, then the efficient
routing combined with effective betweenness based node capability
scheme, abbreviated as (EBC, EFR), can achieve good balance between
the network transmission capacity and designing cost in most cases,
except for networks with rigid hierarchical structures such as HOT
network. In addition, (EBC, EFR) also has good scalability for both
$R_c$ and $C_{max}$. Among all networks, ER network is a markedly
good candidate to achieve cost-effective designing,  especially when
routing algorithm and node capability scheme can not be determined
beforehand.

\acknowledgments This work is partly supported by the National
Natural Science Foundation of China under Grant No. 60673168 and the
Hi-Tech Research and Development Program of China under Grant No.
2008AA01Z203.

\end{document}